# Fluctuating exchange interactions enable quintet multiexciton formation in singlet fission

Miles I. Collins,[1,2)] Dane R. McCamey,[1a)] and Murad J.Y. Tayebjee[2a)]

[1] ARC Centre of Excellence in Exciton Science, School of Physics, University of New South Wales, Sydney, New South Wales 2052, Australia.
[2] School of Photovoltaic and Renewable Energy Engineering, University of New South Wales, Sydney, New South Wales 2052, Australia.
[a)] To whom correspondence should be addressed: dane.mccamey@unsw.edu.au; m.tayebjee@unsw.edu.au

**Abstract**

Several recent electron spin resonance studies have observed a quintet multiexciton state during the singlet fission process. Here we provide a general theoretical explanation for the generation of this state by invoking a time-varying exchange coupling between pairs of triplet excitons, and subsequently solving the relevant time-varying spin Hamiltonian for a range of transition times. We simulate experimental ESR spectra and draw qualitative conclusions about the adiabatic/diabatic transition between triplet pair spin states.

## I. INTRODUCTION

Singlet fission (SF) is a complex photophysical process in which an optically excited singlet exciton transitions to two triplet excitons, each with approximately half the energy of the initial excitation. It is a process which is of broad interest. As a fundamental photophysical process it provides insight into the complex role that spin plays in the dynamics of molecular excitons. Technologically, it has potentially transformational applications across a diverse range of areas. In photovoltaic energy generation, for example, SF provides a potential route to circumvent the Shockley-Queisser limit[1], by allowing wavelengths above the band gap to be more efficiently harvested[2,3]. It can also be used to improve photocatalysis[4] with possible biomedical applications, and there have been suggestions that it may have a role to play in information processing.

Whilst the initial and final states of SF are fairly simple and well understood, the dynamic pathways that allow transitions between them are complex[5–7]. Recent experimental results have revealed the presence of quintet excitons involved in the SF process[8–10], an observation which has become even more interesting as these higher spin states are observed in more classes of molecular system which undergo SF[11–13]. This has raised questions related to the underlying mechanism which generate quintet states, and motivates the work described below.

Existing models for SF[5,14–20] tend to take an electron configuration approach to the description of SF which provides a good description of the stationary states, but has challenges when used to investigate the mechanisms which enable transitions between these states. The ability to transition between spin eigenstates inherently requires a time varying component in the Hamiltonian which describes the system and, as such, a model more suited to this is needed.

In this work we consider variations in the exchange coupling between the two excitons involved in SF as a mechanism for driving transitions between spin manifolds. This approach is based on a somewhat specific and naïve picture – that thermal fluctuations which perturb the planar nature of acene based dimers[11,21–27], particularly those with small bridges[11], are likely to lead to changes in the exchange coupling between the two monomers. Whilst this picture motivates the development of the model below, it is important to note that the approach we take is general for any system where exchange coupling varies in time. However, we focus here on dimer systems, as these have been shown experimentally to have strong quintet populations, and their inherent morphology restrictions significantly reduce the computational degrees of freedom.

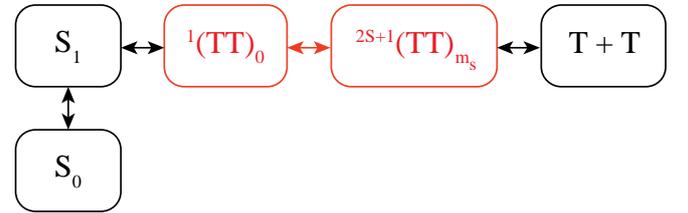

Figure 1: A simplified schematic of the singlet fission process. The transition and states of interest in the present work are highlighted in red.

## II. METHODS

### Hamiltonian

Here we consider two triplets, *a* and *b* in a static magnetic field, ***B***. We first define their orientation using a combination of two rotation matrices:

$$R_{\text{rel}}(\boldsymbol{\varphi}_{a,b}) = R_{\text{rel},z}(\gamma_{a,b})R_{\text{rel},y}(\beta_{a,b})R_{\text{rel},z}(\alpha_{a,b}) \quad (1)$$

$$R_B(\boldsymbol{\varphi}) = R_{B,z}(\gamma_B)R_{B,y}(\beta_B)R_{B,z}(\alpha_B) \quad (2)$$

$$R_{a,b} = R_{\text{rel}}(\boldsymbol{\varphi}_{a,b})R_B(\boldsymbol{\varphi}). \quad (3)$$

Here $R_{\text{rel}}$ describes the relative orientation of the two triplets and $R_B$ describes the orientation of the pair with respect to the applied magnetic field. Both $R_{\text{rel}}$ and $R_B$ are parameterized by three Euler angles using the *y* convention. These rotations are applied to the diagonal representations of the **g** and **D** tensors,

$$\boldsymbol{g}^{a,b} = R_{a,b}\boldsymbol{g_0}R_{a,b}^T, \text{ and} \quad (4)$$

$$\boldsymbol{D}^{a,b} = R_{a,b}\boldsymbol{D_0}R_{a,b}^T. \quad (5)$$

Where $\boldsymbol{g_0}$ and $\boldsymbol{D_0}$ are diagonal representations,

$$\boldsymbol{g_0} = \begin{pmatrix} g_x & 0 & 0 \\ 0 & g_y & 0 \\ 0 & 0 & g_z \end{pmatrix}, \text{ and} \quad (6)$$

$$\boldsymbol{D_0} = \begin{pmatrix} -\frac{1}{3}D + E & 0 & 0 \\ 0 & -\frac{1}{3}D - E & 0 \\ 0 & 0 & \frac{2}{3}D \end{pmatrix} \quad (7)$$

The total spin Hamiltonian is the sum of the Zeeman, zero-field splitting and inter-triplet interaction Hamiltonians:[8]

$$\hat{H} = \hat{H}_z + \hat{H}_{zfs} + \hat{H}_{ab}. \quad (8)$$

Using standard notation for spin operators, tensors and physical constants, we have the individual triplet Hamiltonians,

$$\hat{H}_z^{a,b} = \mu_B \sum_{i,j=x,y,z} B_i g_{ij}^{a,b} \hat{S}_j^{a,b}, \text{ and} \quad (9)$$

$$\hat{H}_{zfs}^{a,b} = \sum_{i,j=x,y,z} \hat{S}_i^{a,b} D_{i,j}^{a,b} \hat{S}_j^{a,b}. \quad (10)$$

The Hamiltonians in the two-triplet basis is

$$\hat{H}_z = \hat{H}_z^a \otimes I + I \otimes \hat{H}_z^b, \text{ and} \quad (11)$$

$$\hat{H}_{zfs} = \hat{H}_{zfs}^a \otimes I + I \otimes \hat{H}_{zfs}^b. \quad (12)$$

The inter-triplet interaction is given by

$$\hat{H}_{ab}(t) = \sum_{i,j=x,y,z} J_{ij}(t)\hat{S}_i^a \otimes \hat{S}_j^b \quad (13)$$

In this work we ignore the dipolar and spin-orbit coupling terms of $J$, such that

$$J_{ij}(t) = \begin{cases} J_{iso} & i = j \\ 0 & i \neq j \end{cases}. \quad (14)$$

It is useful to define the pure spin-state solutions[28] which are solutions to both $\hat{S}^2$ and $\hat{S}_z$:

$$^1(TT)_0 = \frac{1}{\sqrt{3}}(|00\rangle - |-+\rangle - |+-\rangle) \quad (15)$$

$$^3(TT)_0 = \frac{1}{\sqrt{2}}(|-+\rangle - |+-\rangle) \quad (16)$$

$$^5(TT)_0 = \frac{1}{\sqrt{6}}(2|00\rangle + |-+\rangle + |+-\rangle) \quad (17)$$

$$^3(TT)_{\pm 1} = \frac{1}{\sqrt{2}}(|\pm 0\rangle - |0\pm\rangle) \quad (18)$$

$$^5(TT)_{\pm 1} = \frac{1}{\sqrt{2}}(|\pm 0\rangle + |0\pm\rangle) \quad (19)$$

$$^5(TT)_{\pm 2} = |\pm\pm\rangle \quad (20)$$

By examining the commutation relations $[\hat{S}^2, \hat{H}_z] = 0$, $[\hat{S}^2, \hat{H}_{ab}] = 0$, and $[\hat{S}^2, \hat{H}_{zfs}] \neq 0$ we note that under high exchange ($J_{iso} \gg D$), $\hat{H}_{zfs}$ becomes negligible and the eigenstates of $\hat{H}$ are minor perturbations of Equations 13-18. For simplicity, we will refer to the eigenstates of $\hat{H}$ in the high-coupling regime using the notation in Eqs 1.

However, when $|J_{iso}| \lesssim |D|$, $[\hat{S}^2, \hat{H}] \neq 0$ and state-mixing via zero-field splitting occurs. In particular, the $^1(TT)_0$ and $^5(TT)_0$ states are mixed to yield SQ and QS. Therefore it has been suggested that the magnitude of $J_{iso}$, increases from 0 to a maximum value $J_{max} \gg D$ during the ESR experiment.[9]

## Implementation

As shown in Figure 2, we model the change in exchange coupling using two parameters: the transition rate, $1/\tau_r$, and the statistical distributions of rates, $1/\tau_s$. The former describes rate at which the exchange coupling changes during nuclear reorganization, and is expected to be in the range of 100fs – 10 ns.[29] The latter describes the distribution of the activation times, $t_0$, of the nuclear reorganization.

We solve the time-dependent Schrodinger equation,

$$i\hbar \frac{d}{dt}|\Psi(t)\rangle = \hat{H}(t)|\Psi(t)\rangle \quad (21)$$

where $\Psi(0) = {}^1(TT)_0$. The time-dependent term in the Hamiltonian is $\hat{H}_{ab}(t)$, as shown in Figure 2. This is implemented using the Lindblad Master Equation solver in QuTiP (version 4.3.1)[30].

We simulate a time varying exchange interaction

$$J_{iso} = \begin{cases} 0 & t < t_0 \\ J_{max}\left(1 - exp\frac{-(t-t_0)}{\tau_r}\right) & t \geq t_0 \end{cases}. \quad (22)$$

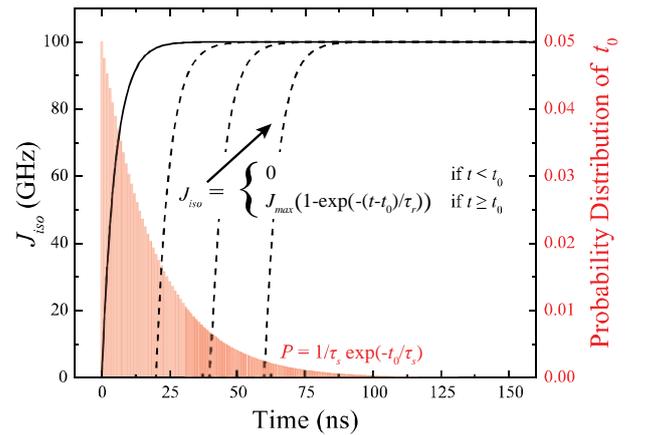

Figure 2: The two transition time constants of interest, $\tau_r$ and $\tau_s$. The former describes the rate of change of the isotropic exchange coupling. The latter describes the statistical distribution of $t_0$.

We repeat this for a range of orientations of the triplet pair relative to the applied static field. Once $J_{iso} = J_{max}$, ($t = t_0 + 5\tau_r$) the state (now stationary) is projected onto the high-exchange eigenstates, which are slight perturbations of eqns 15-20. This yields the population density $\rho'(\boldsymbol{B}, \tau_r, t_0)$.

The spectra at the end of the time evolutions, $s'(|\boldsymbol{B}|, \tau_r, t_0)$, are generated using the `resfields` function in EasySpin 5.2.25[31,32] and a methodology that we have previously described.[8] We use the probability distribution function (as shown in figure 2),

$$P(t_0) = 1/\tau_s \exp\left(\frac{-t_0}{\tau_s}\right)$$

to generate a single powder spectrum depending on $\tau_r$ and $\tau_s$

$$s(|\boldsymbol{B}|, \tau_r, \tau_s) = \sum_i s'(|\boldsymbol{B}|, \tau_r, t_{0,i}) P_i(t_0). \quad (23)$$

Similarly, the populations of each of the nine spin states in Eqs. 7-12 are given by

$$\rho(\boldsymbol{B}, \tau_r, \tau_s) = \sum_i \rho'(\boldsymbol{B}, \tau_r, t_{0,i}) P_i(t_0). \quad (24)$$

Since a goal of this work is to provide a framework for simulating time-resolved ESR spectra we limit the characteristic lifetimes, $\tau_r$ and $\tau_s$, to being below the time resolution of X-band microwave cavities (~50 ns).

## III. RESULTS

In the calculations below we model two triplets using the semi-empirical values of the pentacene dimer system **BP3**, as shown in Figure 3.[8] Since this is a symmetric dimer, the $\boldsymbol{g}$ and $\boldsymbol{D}$ tensors are identical. The parameters are given in Table 1. This approach described in the preceding section is general and, as such, any dimeric system can be simulated by changing the parameters in Table 1.

Table 1: Parameters.

| Parameter | Value |
|---|---|
| $g_x = g_y = g_z$ | 2.002 |
| $D$ | 1138 MHz |
| $E$ | 19 MHz |
| $\boldsymbol{\varphi_a}$ | [-60°, 2.39°, 60°] |
| $\boldsymbol{\varphi_b}$ | [-60°, -2.39°, 60°] |
| $J_{max}$ | 100 GHz |
| $J_{iso}(0)$ | 0 |

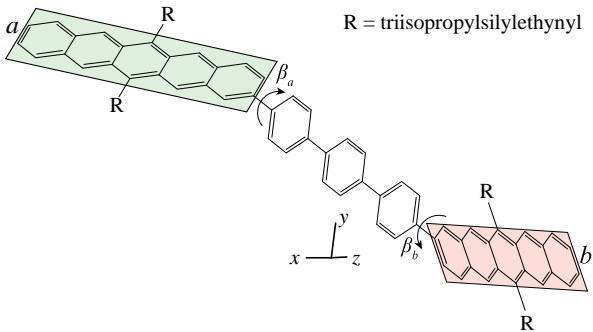

Figure 3: The bipentacene, **BP3**, contains the two triplets, *a* and *b* which are rotated by $R_{rel}$ with respect to the molecular Cartesian axes. Rotations about the triphenyl bridge are exaggerated.

Figure 4 shows the populations of each spin state as a function of $t_0$ for a magnetic field applied along the three principle axes. The majority of the population density oscillates between $^5(TT)_0$ and $^1(TT)_0$ with regular periods. The period of these oscillations can be derived from the eigenvalues of $\hat{H}_{zfs}$; in symmetric triplets the SQ and QS states are separated by approximately $D \pm 3E$ for $\boldsymbol{B}||y,x$ and $2D$ for $\boldsymbol{B}||z$; this gives rise oscillations of approximately $2\pi/(D \pm E)$~5.5ns and $\pi/D$ ~ 2.8ns; deviations from these values arise from the angle between the planes of the two chromophores.

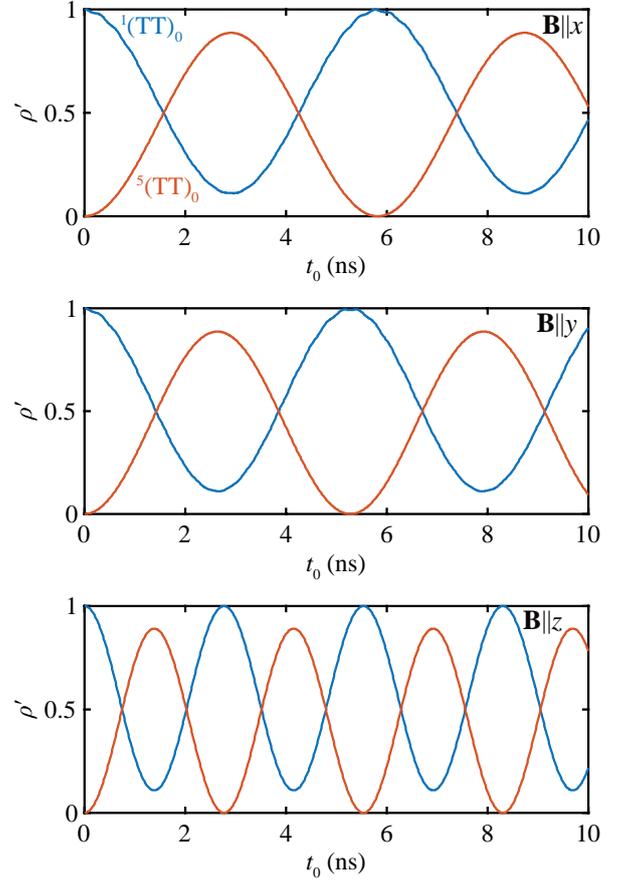

Figure 4: Projection upon the $^1(TT)_0$ and $^5(TT)_0$ states as a function of time where $\tau_r = 100$fs for $|\boldsymbol{B}| = 350$mT in the *x*, *y* and *z* directions.

Figure 5(a) and the animations in the Supporting Information show a clear orientation dependence for the generation of the $^5(TT)_0$ state. For statistical rates of the order of the oscillations in Figure 4, significant $^5(TT)_0$ character is generated. Interestingly, when $\theta$ is close to the magic angle (~54.7°) less $^5(TT)_0$ is observed. This is because the dipole-dipole interaction is reduced and the energy splitting between the SQ and QS states approaches zero, giving rise to slow mixing. This is explicitly shown in Figure 6, where the blue line corresponds to the solutions of $\hat{H}$ when $\theta = 54.7°$ and $\phi = 0$; in this case the SQ and QS states are near-degenerate.

Figure 5(b) shows the population distribution of triplet pair states for a slow transition lifetime (50 ns). Here the distribution of pair states is very different to (a). The $^5(TT)_{-2}$ and $^5(TT)_{-1}$ have larger population densities than the $^5(TT)_0$ state. There is also a clear orientation dependence of the generation of these states. For instance, the $^5(TT)_{-2}$ state is only generated when $\theta$ is close to 90°.

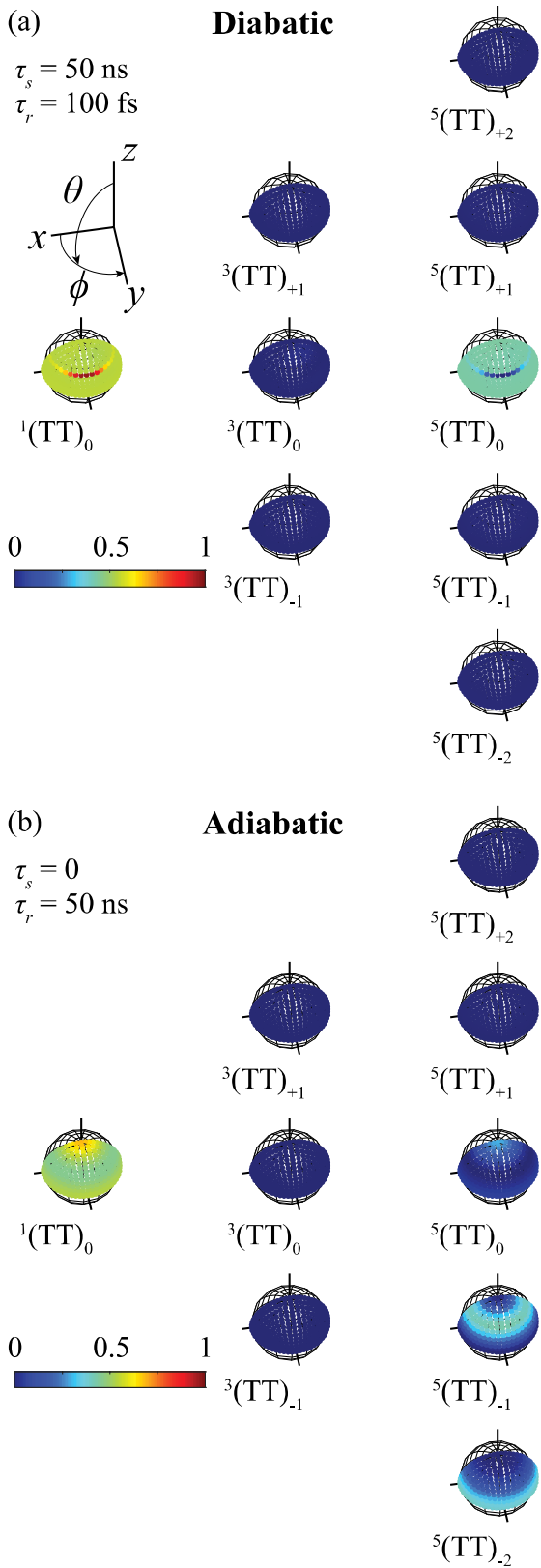

Figure 5: Orientation dependence of the population densities of for an (a) (multimedia view, varying $\tau_s$) diabatic and an (b) (multimedia view, varying $\tau_r$) adiabatic transition. See Supporting Information for versions of these images on logarithmic scales. Inset: Definition of polar coordinates.

This is explained by examining Figure 6. When the applied field is aligned along the *x*-axis of the molecule (black lines) there is an avoided crossing between the $^5(TT)_{-2}$ and $^1(TT)_0$ states; however, no avoided crossing is observed between the $^5(TT)_{-1}$ and $^1(TT)_0$ states. When $\theta = 54.7°$ (blue lines) avoided crossings exist between $^1(TT)_0$ and both $^5(TT)_{-2}$ and $^5(TT)_{-1}$, and indeed both are populated (cf. Fig 5(b)). If the applied field is applied along the molecular *z*-axis (black lines), only the initial mixing between the SQ and QS states may give rise to population density in the $^5(TT)_0$ state.

Figure 7(a) and (b) respectively show the population density as a function of the statistical and transition lifetimes in the diabatic and adiabatic cases. In Figure 7(a) the population densities of the $^1(TT)_0$ and $^5(TT)_0$ approach each other and then plateau as $\tau_s$ is increased. Since the singlet and quintet spin states are even with respect to exchange and the triplet states are odd, we do not expect any population density in the triplet manifold.[28] However, the chromophores are not perfectly aligned ($\varphi_{a,b} \neq 0$), and this parity argument for breaks down. This gives rise to non-zero population density in the $^3(TT)_0$ state.

The corresponding data in Figure 7(b) is more complex. As shown in Figure 6, when the value of $J_{iso}$ is increased mixing can occur between the $^1(TT)_0$ and the lower lying quintet states. If the rate of change of $J_{iso}$ is low, then state-mixing occurs giving rise to populations in the quintet manifold. Some $^3(TT)_x$ population density is also observed at longer transition lifetimes.

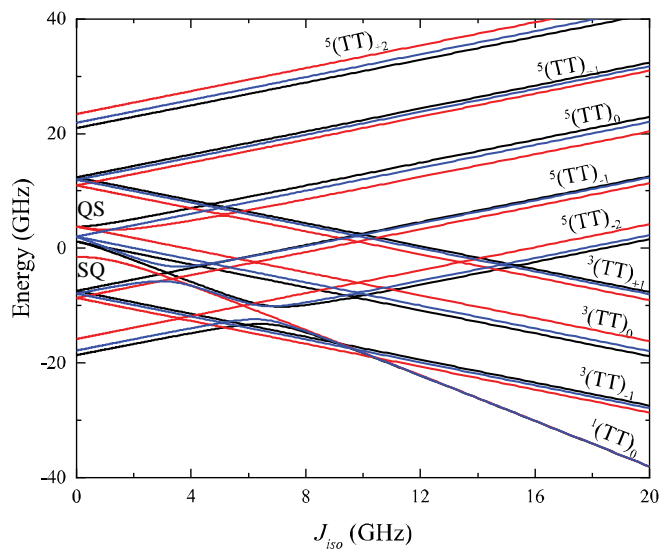

Figure 6: The effect of $J_{iso}$ on the energy levels of the triplet pair Hamiltonian, with a magnetic field strength of 350 mT. The black, red, and blue lines correspond to a magnetic field aligned along the *x*, *z*, and [$\theta$=54.7°, $\phi = 0$] molecular directions.

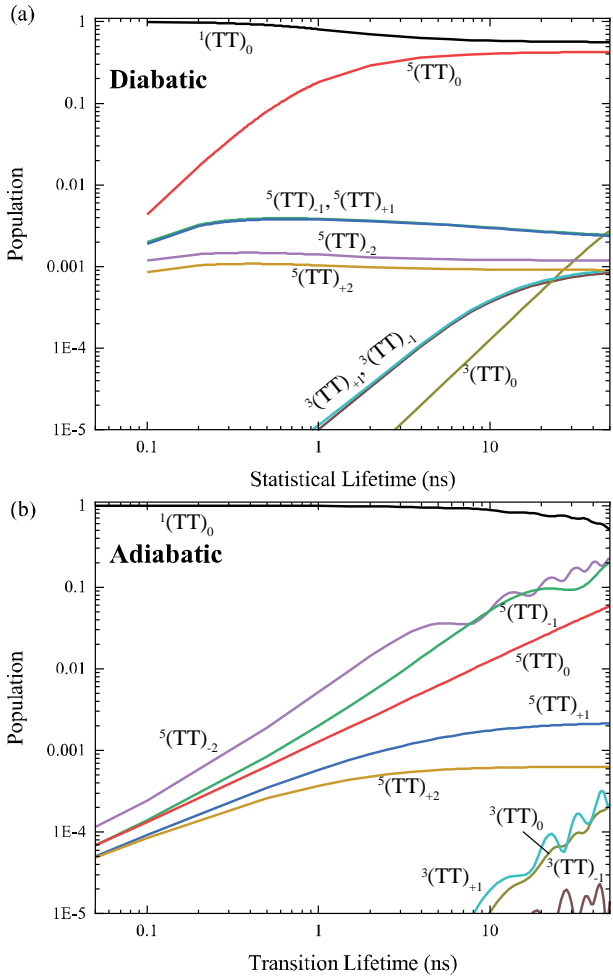

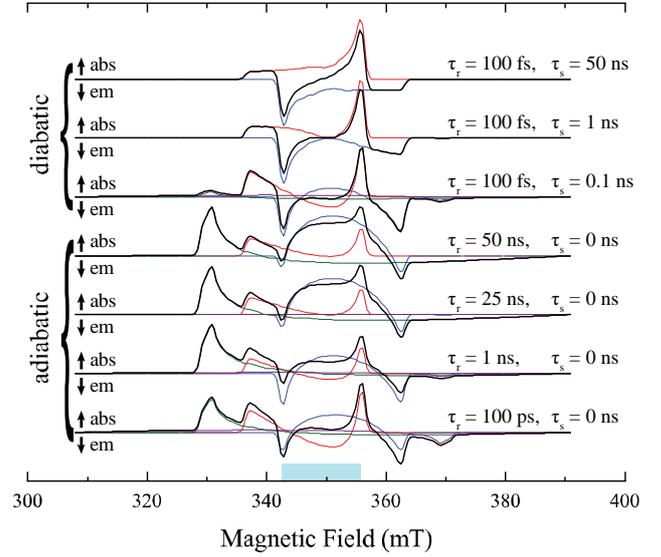

Figure 7: Population densities averaged over all magnetic field orientations as a function of the (a) statistical lifetime, $\tau_s$ (where $\tau_r = 100$ fs) and (b) transition lifetime, $\tau_r$ (where $\tau_s=0$).

## IV. DISCUSSION

Now that we have established a possible route to the quintet manifold it is useful to simulate the corresponding ESR spectra and compare them to recent experimental reports.

Figure 8 shows the simulated spectra for a series of values of $\tau_r$ and $\tau_s$. To centre the data around 350 mT we use a microwave frequency of $\nu = 9.807$ GHz. We stress in the following discussion, that we have only carried out calculations for pentacene chromophores that are approximately aligned; nevertheless, qualitative comparisons may be drawn between our calculations and the experimental ESR spectra of other systems.

The first two reports of a quintet state observation using ESR were measurements of **BP3** in a toluene matrix and TIPS-tetracene thin films.[8,9] The shape of initial quintet spectra looked very similar to the simulation of a diabatic transition with a long statistical lifetime. The present work accords with the original assertion that $J_{iso}$ transitioned from a small to large value with respect to D;[9] however, we show here that the statistical distribution, $P(t_0)$, has a profound effect on the shape of the ESR spectrum.

Figure 8: Calculated normalized ESR spectra for quintets produced via singlet fission. Red, blue, green, and magenta lines respectively correspond to the $^5(TT)_0 \leftrightarrow {}^5(TT)_{+1}$, $^5(TT)_0 \leftrightarrow {}^5(TT)_{-1}$, $^5(TT)_{-2} \leftrightarrow {}^5(TT)_{-1}$, and $^5(TT)_{+1} \leftrightarrow {}^5(TT)_{+2}$, transitions. The total spectra are given in black lines. The blue shaded box indicates the region of the spectrum where net positive absorption features arising from adiabatic transitions between $^5(TT)_0 \leftrightarrow {}^5(TT)_{-1}$ are prominent.

In the third observation of a quintet state, Basel et al. showed that the singlet to quintet transition in a non-conjugated pentacene dimer was 550 ns at 105 K.[13] The tr-EPR showed a $^5(TT)$ spectrum (Figure 5 in Ref. [13]) which resembles the adiabatic transition, where $\tau_r = 50$ns. Given the trend in Figure 8, we expect the spectrum with $\tau_r = 550$ns to also be similar. We note that the pentacene chromophores are non-parallel. However, the angle between the planes of the chromophores is relatively small, so the qualitative comparison here holds since the effect of $R_{rel}$ on the overall Hamiltonian will also be small.

In another pentacene dimer bridged by the non-conjugated bicyclooctane moiety, we also measured a similarly broad absorptive feature (Figure 5 in Ref. [11]). This supports the hypothesis that non-conjugated bridges in isolated dimers may give rise to more adiabatic transitions between $^1(TT)_0$ and $^{2S+1}(TT)_{ms}$ than their conjugated counterparts. This can be explained by considering that the value of $J_{max}$ will be lower in non-conjugated systems, and therefore the derivative of $J_{iso}$ with respect to time will be small, allowing for adiabatic transitions.

ESR experiments of singlet fission have also repeatedly yielded triplet pair populations which give rise to a net absorptive ESR spectrum, particularly in the range $\frac{h(\nu-D/3)}{\mu_B g} \lesssim |B| \lesssim \frac{h(\nu+D/3)}{\mu_B g}$ (indicated by the blue region on the x-axis in Figure 8). The results herein suggest that this may be due to a adiabatic transition from $^1(TT)_0$ to $^5(TT)_{-1}$ for particular values of $\theta$.

An interesting upshot of this work is that we have assumed that the inter triplet exchange coupling is a function of time.

This is necessarily true for a transition which involves nuclear coordinate rearrangement. However, we note that we have placed a requirement that $J_{\mathrm{iso}}$ *must*, at some point, be small with respect to $D$, and at a later time be large with respect to $D$ (in this case set to 100 GHz). The fact that the $^5(TT)_0$ state is observed in so many pentacene dimer systems[11–13] suggests that this arises from an inherent property of singlet fission; i.e. the initially-generated $^1(TT)_0$ state has a low exchange coupling in these systems.

## V. CONCLUSIONS

We have shown that a time varying exchange interaction in dimeric SF systems drives interconversion between the singlet and quintet pair manifolds. For diabatic transitions with statistical lifetimes on the order of several nanoseconds, quintets are preferentially generated in the $^5(TT)_0$ state, whilst adiabatic transitions provide pathways to access the $^5(TT)_{-1}$ and $^5(TT)_{-2}$ states.

Taking the ensemble average of a statistical distribution of the states which result from such a process reproduces population distributions which are highly consistent with those observed experimentally. Our model uses two extra parameters, $\tau_r$ and $\tau_s$, but grounds the work in a solid theoretical foundation. Further development will include deriving (possibly non-monotonic) $J_{\mathrm{iso}}(t)$ from time-dependent density functional theory.

Finally, we note that the time-varying approach used here can be easily modified to investigate other mechanisms, including variations in both dipolar and hyperfine interactions. Or, indeed, to simulate other experiments such as magnetic-field-dependent photoluminescence.

## VI. SUPPLEMENTARY MATERIAL

Additional animations of figure 5a and 5b on a log scale are in supplementary material.

## VII. ACKNOWLEDGEMENTS

This work was supported by the ARC Centre of Excellence in Exciton Science (CE170100026). M.J.Y.T. thanks Dr Florian Schroeder for useful discussions.